\documentclass[amsmath,
preprint,
endfloats*, aps,prl]{revtex4-1}
\usepackage{graphicx}
\usepackage{dcolumn}
\usepackage[colorlinks=true, pdfstartview=FitV,
linkcolor=blue,citecolor=blue, urlcolor=blue]{hyperref}
\begin{document}
\bibliographystyle{unsrt} \title{Harnessing nuclear spin polarization
  fluctuations in a semiconductor nanowire}
\author{P. Peddibhotla$^1$, F. Xue$^1$, H. I. T. Hauge$^2$,
  S. Assali$^2$, E. P. A. M. Bakkers$^{2,3}$, M. Poggio$^1$}
\affiliation{$^1$Department of Physics, University of Basel, 4056 Basel, Switzerland \\
  $^2$Department of Applied Physics, Eindhoven University of Technology, 5600 MB Eindhoven, The Netherlands \\
  $^3$Kavli Institute of Nanoscience, Delft University of Technology,
  2600 GA Delft, The Netherlands} \date{\today}
\begin{abstract}
  Soon after the first measurements of nuclear magnetic resonance
  (NMR) in a condensed matter system, Bloch \cite{Bloch:1946}
  predicted the presence of statistical fluctuations proportional to
  $1/\sqrt{N}$ in the polarization of an ensemble of $N$ spins.  First
  observed by Sleator et al.\ \cite{Sleator:1985}, so-called ``spin
  noise'' has recently emerged as a critical ingredient in
  nanometer-scale magnetic resonance imaging (nanoMRI)
  \cite{Degen:2009,Mamin:2013,Staudaucher:2013,Nichol:2013}.  This
  prominence is a direct result of MRI resolution improving to better
  than $(100 \text{ nm})^3$, a size-scale in which statistical spin
  fluctuations begin to dominate the polarization dynamics.  We
  demonstrate a technique that creates spin order in nanometer-scale
  ensembles of nuclear spins by harnessing these fluctuations to
  produce polarizations both larger and narrower than the natural
  thermal distribution.  We focus on ensembles containing $\sim 10^6$
  phosphorus and hydrogen spins associated with single InP and GaP
  nanowires (NWs) and their hydrogen-containing adsorbate layers.  We
  monitor, control, and capture fluctuations in the ensemble's spin
  polarization in real-time and store them for extended periods.  This
  selective capture of large polarization fluctuations may provide a
  route for enhancing the weak magnetic signals produced by
  nanometer-scale volumes of nuclear spins.  The scheme may also prove
  useful for initializing the nuclear hyperfine field of electron spin
  qubits in the solid-state.
\end{abstract}

\maketitle

Spin noise is a phenomenon present in all spin ensembles that begins
to exceed the mean thermal polarization as the size of the ensemble
shrinks. These fluctuations have random amplitude and phase and have
been observed in a wide variety of nuclear spin systems including in
liquids by conventional NMR \cite{McCoy:1989,Gueron:1989} and in the
solid-state using a superconducting interference device
(SQUID)\cite{Sleator:1985}, by force-detected magnetic resonance
\cite{Mamin:2003}, or by nitrogen-vacancy (NV) magnetometry
\cite{Mamin:2013,Staudaucher:2013}.  In addition to finding
applications in nanoMRI, spin noise has been used in MRI specially
adapted for the investigation of extremely delicate specimens, in
which external radio frequency (RF) irradiation is not desired
\cite{Muller:2006}.  Through the hyperfine interaction, nuclear spin
noise also sets limits on the coherence of electronic qubits in the
solid state
\cite{Merkulov:2002,Khaetskii:2002,Childress:2006,Kuhlmann:2013}.
Various efforts to mitigate these nuclear field fluctuations by
hyperfine-mediated nuclear spin preparation have been developed, both
in quantum dots (QDs)
\cite{Coish:2004,Reilly:2008,Latta:2009,Vink:2009,Bluhm:2010} and in
NV centers in diamond \cite{Togan:2011}.  We report on a manipulation
and initialization technique which applies to arbitrary
nanometer-scale samples, i.e.\ it does not require specialized
structures providing a controllable electronic spin and a strong
hyperfine interaction.  Rather, our technique requires a detector
sensitive enough to resolve nuclear spin noise and the ability to
apply RF electro-magnetic pulses.

Conventional NMR and MRI techniques rely on manipulating the mean
thermal polarization to produce signals.  Statistical nuclear
polarization fluctuations exceed the mean thermal polarization below a
critical number of spins $N_c = \frac{3}{I (I+1)} \left ( \frac{k_B
    T}{\hbar \gamma B_0} \right )^2$, where $I$ is the spin quantum
number, $k_B$ is the Boltzmann constant, $T$ is the temperature,
$\hbar$ is Planck's constant, $\gamma$ is the gyromagnetic ratio, and
$B_0$ is the magnetic field \cite{Xue:2011}.  For $^1$H spins at $T =
1$ K in an applied magnetic field $B_0 = 1$ T, $N_c = 10^6$, which in
most organic samples corresponds to a critical volume $V_c = (24
\text{ nm})^3$.  Under the ambient conditions of recent NMR
experiments with NV spin sensors, statistical polarization begins to
dominate for even larger samples with $V_c = (6 \text{ }
\mu\text{m})^3$ \cite{Mamin:2013,Staudaucher:2013}.  In order to
measure and initialize such ensembles, it is therefore worthwhile to
consider techniques designed to both detect and control nuclear
polarization fluctuations.  Although nuclear magnetic signals from
such small volumes are weak, if a detector is able to resolve spin
noise in real-time, i.e.\ faster than the spin correlation time
$\tau_m$, large fluctuations can be captured.  Once captured, these
fluctuations can then be used to initialize the polarization of
nanometer-scale spin ensembles with a fixed sign and magnitude.  Such
initialization schemes could provide the basis for enhancing signals
from small samples and for realizing advanced pulse protocols which
can be borrowed directly from conventional NMR.

In 2005, Budakian et al.\ first demonstrated a technique for the
capture and storage of electron spin fluctuations
\cite{Budakian:2005}.  Inspired by this work, we have adapted the
protocol to nanometer-scale ensembles of nuclear spins. Given the
relevance of these ensembles to nanoMRI and to the decoherence of
solid-state qubits, this implementation could find wide
applicability. Furthermore, the longer spin relaxation time of nuclear
spins compared to electron spins -- sometimes as long as several hours
at cryogenic temperatures -- allows captured spin order to be stored
for far longer times.

We study two separate samples: an InP and a GaP NW, both grown with
the vapor-liquid-solid method in a metal-organic vapor-phase epitaxy
reactor using gold droplets as catalyst \cite{Assali:2013}.  Using a
magnetic resonance force microscope (MRFM), we measure the
polarization of nanometer-scale ensembles of $^{31}$P nuclei within
each NW and of $^1$H nuclei contained in the hydrocarbon adsorbate
layer on the surface.  Adiabatic rapid passage (ARP) pulses of the
transverse field \cite{Tannus:1996} are used to cyclically invert the
polarization of nanometer-scale volumes of a particular nuclear
isotope.  In a magnetic field gradient, these periodic inversions
generate an alternating force that drives the mechanical resonance of
an ultrasensitive cantilever; the resulting oscillations, which are
proportional to this force, are detected by a fiber-optic
interferometer.

We feed the cantilever force signal to a lock-in amplifier referenced
to the periodic spin inversions and monitor its in-phase ($X$) and
quadrature ($Y$) amplitudes. The limiting source of noise in the
measurement is the thermal noise of the cantilever, which has a random
phase and on average contributes equally to $X$ and $Y$.  On the other
hand, force fluctuations due to the nuclear spin polarization are in
phase with the inversion pulses and contribute only to $X$.  Thus, as
shown in Fig.~\ref{fig2}a, we can monitor $Y(t)$, the thermal force,
and $X(t)$, the thermal force plus the force due to the nuclear spin
inversions.  $X(t)$ is dominated by the large fluctuations and the
long $\tau_m$ of the spin noise, while the thermal noise in $Y(t)$ has
a smaller amplitude and a shorter correlation time set by the damped
cantilever force sensor (see methods).  $\tau_m$ is limited by the
magneto-mechanical noise originating from the thermal motion of the
cantilever in a magnetic field gradient and by the ARP pulse
parameters \cite{Degen:2008}.  Since the contribution of the spin
signal to $X(t)$ is large enough, we can follow -- in real-time -- the
instantaneous nuclear spin imbalance in the rotating frame.
Fig.~\ref{fig2}b shows the variances $\sigma_X^2$ and $\sigma_Y^2$,
which give the variance due only to the thermal noise $\sigma_{T}^2 =
\sigma_Y^2$ and the variance due only to the spin noise, $\sigma_{S}^2
= \sigma_X^2 - \sigma_Y^2$.  $\sigma_{S}^2$ is in turn related to the
number of nuclear spins in the detection volume (see methods).  The
typical size of the spin ensembles we measure is between $2 \times
10^5$ and $7 \times 10^5$ for $^1$H and between $6 \times 10^5$ and $1
\times 10^7$ for $^{31}$P.  Given the density of typical adsorbate
layers and InP and GaP, the detection volumes discussed here are
between $(13 \text{ nm})^3$ and $(21 \text{ nm})^3$ for $^1$H and
between $(30 \text{ nm})^3$ and $(80 \text{ nm})^3$ for $^{31}$P.

The real-time measurement of spin noise allows us to react to and
control the fluctuations in polarization \cite{Budakian:2005}.  As a
demonstration, we conditionally apply RF $\pi$-inversion pulses to
both rectify and narrow the naturally occurring polarization
distribution.  This feedback is realized via a field-programmable gate
array (FPGA) in conjunction with an arbitrary waveform generator
(AWG).  As shown in Fig.~\ref{fig3}a, when $X(t)$ exceeds a
predetermined threshold, the protocol applies a $\pi$-inversion, by
inserting an ARP pulse with a duration equivalent to a full cantilever
cycle rather than the usual half-cycle.  As a result, the spin
ensemble's periodic inversion at the cantilever frequency undergoes a
$180^{\circ}$ phase shift.  In this way, we can either rectify or
narrow $X(t)$ by setting appropriate thresholds as shown in
Fig.~\ref{fig3}b.  Since only fluctuations due to nuclear spins are
affected by the $\pi$-inversions, the effectiveness of the protocol
depends on the fraction of $X(t)$ arising from spin compared to
thermal fluctuations, i.e.\ the larger the power signal-to-noise ratio
(SNR) $\sigma_{S}^2/\sigma_{T}^2$ is, the more effective the control
of $X(t)$ will be.  As shown in Fig.~\ref{fig3}c, this process can
produce both hyper-polarized and narrowed nuclear spin distributions
in the rotating frame.

In addition to controlling an ensemble's natural spin fluctuations, we
can also capture especially large fluctuations \cite{Budakian:2005}.
As shown in Fig.~\ref{fig4}, using the FPGA and AWG, we continuously
measure the rotating-frame spin fluctuations by monitoring $X(t)$.
Once $X(t)$ reaches a predetermined threshold $X_c$, the ARP pulses
are tuned out of resonance with the nuclear spin ensemble, leaving the
instantaneous spin polarization pointing along $\mathbf{B}_0$ and
transferring the transient spin order to the laboratory frame.  Since
the spin ensemble does not respond to the off-resonant ARP pulses, the
cantilever is no longer driven by spin forces.  In this way,
hyper-polarized states of the nuclear spin ensemble can be captured
and stored in the laboratory frame for as long as $T_1$.  Crucially,
the time-scale required to capture the spin order -- here given by the
inverse of the cantilever frequency $1/f_c$ -- must be much shorter
than the rotating-frame correlation time $\tau_m$.  To confirm that
the nuclear polarization has in fact been stored, we can reapply the
resonant ARP pulses after some storage time $T_{\text{store}}$ to
readout the polarization, once again cyclically inverting the
polarization.  These spin inversions again drive the cantilever
motion, and the amplitude of $X(t)$ reflects the size of the retrieved
fluctuation.

In an idealized case, in which the spin component of the captured
fluctuation $X_c$ is fully projected onto $\mathbf{B}_0$, the stored
fluctuation $X_s$ will be normally distributed with a mean $\left <
  X_s \right > = X_c \frac{\sigma_S^2}{\sigma_S^2 + \sigma_T^2}$ and a
variance $\sigma_{X_s}^2 = \frac{\sigma_S^2 \sigma_T^2}{\sigma_S^2 +
  \sigma_T^2}$ (see supplementary material).  The fact that $\left <
  X_s \right > < X_c$ reflects the finite SNR of the measurement, in
this case limited by the cantilever's thermal noise.  Note also that
the distribution of the stored polarization is narrowed, i.e.\ it has
a reduced variance, compared to the variance of $X(t)$ under normal
evolution ($\sigma_{X_s}^2 < \sigma_X^2 = \sigma_S^2 + \sigma_T^2$).
In the limit of large SNR ($\sigma_S^2 \gg \sigma_T^2$), $\left < X_s
\right > \rightarrow X_c$ and $\sigma_{X_s}^2 \rightarrow 0$.  If
during $T_{\text{store}}$ this stored polarization undergoes
negligible relaxation in the laboratory frame, the corresponding
retrieved fluctuation $X_r$ has a mean $\left < X_r \right > = \left
  <X_s \right >$ and a variance $\sigma_{X_r}^2 = \sigma_{X_s}^2 +
\sigma_{T}^2 < \sigma_X^2$.  Note that $\sigma_{X_r}^2 >
\sigma_{X_s}^2$ due to the finite SNR of the retrieval measurement.

In order to compare our experiments to this idealized case, we measure
$\left < X_r \right >$ and $\sigma_{X_r}^2$ for nanometer-scale
$^{31}$P and $^1$H spin ensembles.  As shown in Fig.~\ref{fig5},
$\left < X_r \right >$ and $\tau_m$ are extracted from bi-exponential
fits to $X(t)$ during the readout sequence using our knowledge of the
lock-in time constant $\tau_l$.  Deviations of $\left < X_r \right >$
from $\left < X_s \right >$ could be caused by the spin-lattice
relaxation of the polarization in the laboratory frame -- set by $T_1$
-- or by the incomplete projection of the polarization onto
$\mathbf{B}_0$.  For both $^{31}$P and $^1$H in Fig.~\ref{fig5}a and
b, these deviations are negligible within our error for two values of
$T_{\text{store}}$.  Limitations of the experimental hardware
prevented measurements for larger $T_{\text{store}}$, although our
data show that $T_1 \gg 20$ s for $^{31}$P in InP and $T_1 \gg 2.5$ s
for $^1$H on GaP at $4.2$ K.

Fig.~\ref{fig4}b shows the reduced variance of the retrieved
polarization $\sigma_{X_r}^2$, eventually approaching $\sigma_X^2$
after a time on the order of $\tau_m$ in the rotating frame.  Note
that $\tau_l$ is much shorter than the time over which
$\sigma_{X_r}^2$ evolves, excluding the lock-in as a source of the
behavior.  This result demonstrates that the distribution of the
captured fluctuations is indeed narrowed relative to the natural
distribution of the nuclear spin fluctuations.  The ability to prepare
such distributions may find application in quantum information
processing with solid-state electron spins, which is often limited by
the random nuclear field distribution in the host material
\cite{Coish:2004,Reilly:2008,Latta:2009,Vink:2009,Bluhm:2010}.  The
nuclear spin ensemble in such a system could be initialized before
each measurement by a scheme based on the capture of random
polarization fluctuations, thus enhancing the electron-spin dephasing
time.  The degree of narrowing demonstrated in Fig.~\ref{fig4}b
represents a factor of $2.5$ compared to the natural distribution and
is limited by the SNR of the measurement.

The size of the captured spin fluctuation is, in principle, only
limited by the amount of time one is willing to wait during the
capture step.  For the normally distributed random variable $X(t)$,
the average amount of wait time required to capture a fluctuation
$X_c$ is given by $T_{\text{wait}} = \frac{2}{n_0} e^{X_c^2 / (2
  \sigma_X^2)}$, where $n_0$ is the average number of times $X(t)$
crosses zero per second (see supplementary material) \cite{Rice:1944}.
For example, given that $n_0 = 0.2$ Hz is a typical value in our
experiments, fluctuations of $3 \sigma_X$ are expected after just
$T_{\text{wait}} = 15$ min, or alternatively for $T_{\text{wait}} = 1$
hr, a fluctuation of $3.4 \sigma_X$ can be expected.  For a $^{31}$P
spin ensemble with $N = 10^6$ at $B = 6$ T and $T = 4.2$ K, the
standard deviation of the statistical polarization fluctuations is
given by $\rho_S = \sqrt{\frac{I+1}{3I} \frac{1}{N}} = 0.1\%$ and its
mean thermal polarization is $\rho_B = \frac{I+1}{3} \frac{\hbar\gamma
  B}{k_B T} = 0.06\%$ \cite{Xue:2011}.  Therefore, in the limit of
large SNR where $\sigma_X$ is dominated by spin fluctuations, we can
expect to capture polarizations of $4.8 \rho_B$ in 15 min and $5.5
\rho_B$ in 1 hr.  The polarization captured in a given
$T_{\text{wait}}$ could be increased even further by reducing $\tau_m$
and therefore increasing $n_0$, e.g.\ through the periodic
randomization of the spin ensemble using bursts of
$\frac{\pi}{2}$-pulses \cite{Degen:2007}.  In principle, the nuclear
spin decoherence time $T_2$ sets the lower limit for $\tau_m$.  As the
size of the spin ensemble shrinks, the size of the achievable
polarization increases as $1/\sqrt{N}$, making such a protocol
increasingly relevant as detection volumes continue to shrink.  Given
that conventional pulse protocols based on thermal polarization
require an initialization step taking at least $T_1$, waiting for an
extended time to capture a large spin fluctuation may be attractive --
especially when the magnitude of the captured fluctuation greatly
exceeds the possible thermal polarization.

Note that the crucial step in the capture protocol is that the
monitoring and capture occurs in the rotating frame, where the time
between statistically independent spin configurations -- set by
$\tau_m$ -- is much faster than the equivalent time $T_1$ in the
laboratory frame.  This reduced rotating-frame correlation time,
allows the system to quickly explore its spin configuration space.  On
the other hand, once a large fluctuation is captured and transferred
to the laboratory frame, the long $T_1$ effectively freezes the
ensemble in this rare configuration.

We therefore show the long-term storage of large polarization
fluctuations arising from spin noise in nanometer-scale nuclear spin
ensembles with $N \sim 10^6$.  Storage times as long as 20 s for
$^{31}$P are demonstrated, limited only by the measurement hardware.
The ultimate limit to these times is set only by $T_1$ in the
laboratory frame, which at low temperatures is typically extremely
long, exceeding hours for some nuclear spins.  While these results
were obtained with a low-temperature MRFM, the capture and storage of
spin fluctuations should be generally applicable to any technique
capable of detecting and addressing nanometer-scale volumes of nuclear
spins in real-time.

The ability to initialize the polarization of a small nuclear spin
ensemble is important for the development of future nanometer-scale
NMR techniques or possibly for the implementation of solid-state
qubits and nuclear spin memory devices.  Given that the mean thermal
polarization is typically negligible in nanometer-scale samples, other
methods to create nuclear spin order must be considered.  When
polarization cannot be created via effects such as Overhauser-mediated
dynamic nuclear polarization or electron-nuclear double resonance, the
selective capture of large statistical fluctuations provides a viable
alternative.  The ensembles polarized in this work, nanometer-scale
volumes of $^1$H on an adsorbate layer and $^{31}$P in a single
semiconducting NW, demonstrate the types of samples which could
benefit from this technique.  One could imagine, for instance, such
nuclear polarization capture processes enhancing the weak MRI signals
of a nanometer-scale $^1$H-containing biological sample on a surface
or of a semiconducting nanostructure.

\section{Methods}

\begin{figure}[t]
        \includegraphics[width=0.66\textwidth]{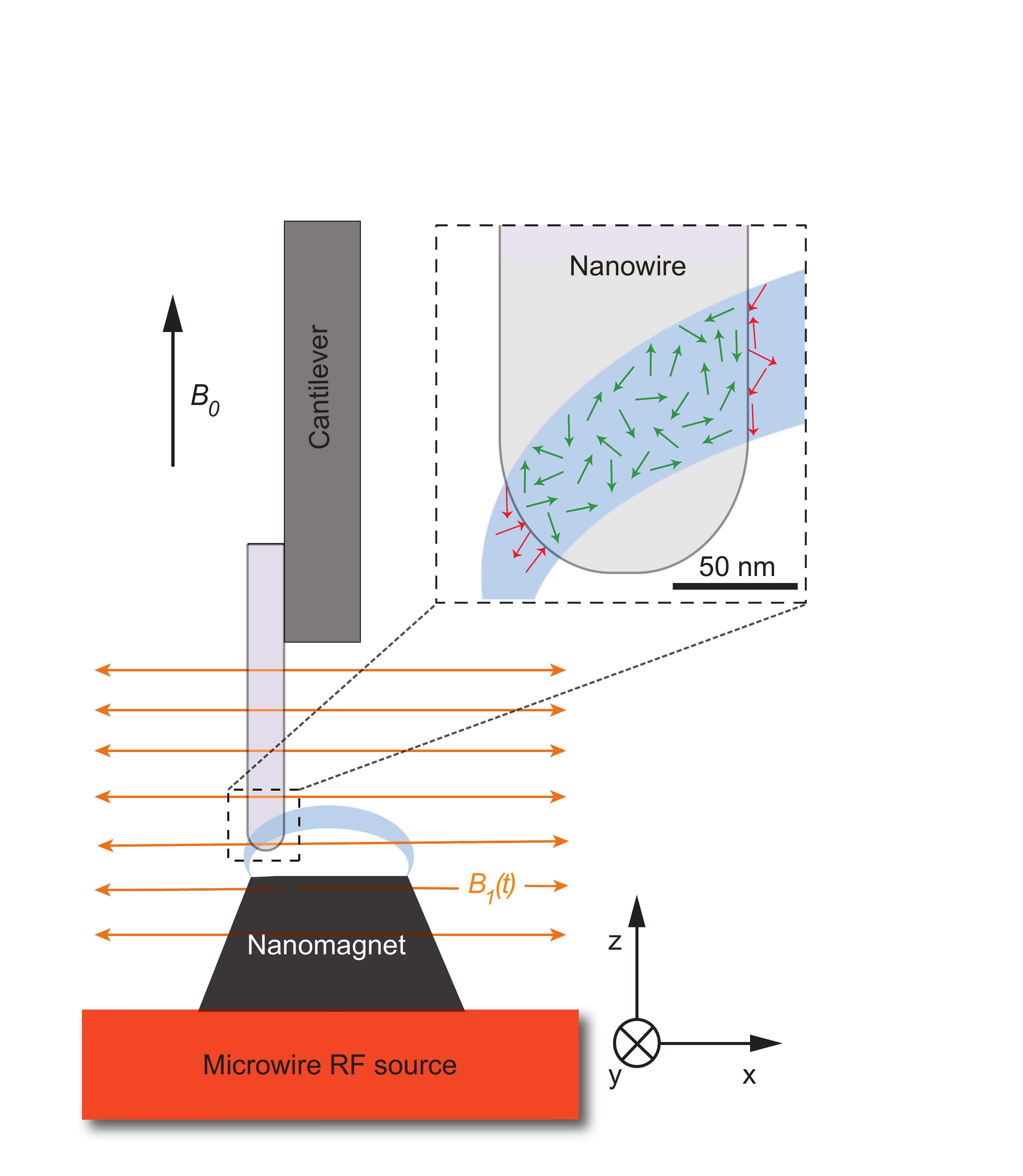}
        \caption{{\bf Schematic of the experimental geometry.}  The
          end of the NW, which is affixed to an ultrasensitive
          cantilever, is positioned 100 nm away from the nanomagnet.
          Below the tip, a microwire RF source generates ARP pulses to
          invert the nuclear spin ensemble within a nanometer-scale
          ``resonant slice'' (in light blue).  Two types of spin
          ensembles are investigated: one composed of $^{31}$P nuclei
          within the NW lattice (green spins) and another consisting
          of $^1$H nuclei from the thin adsorbate layer on the NW
          surface (red spins).}
  \label{fig1}
\end{figure}

\begin{figure}[t]
	\includegraphics[width=\textwidth]{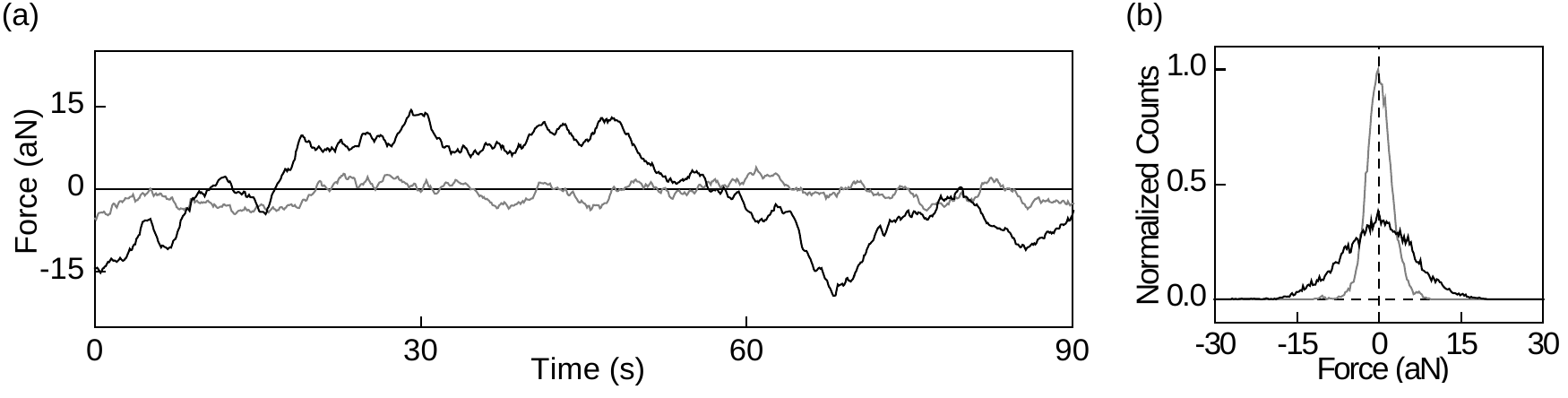}
        \caption{{\bf Spin noise from an ensemble of $6 \times 10^5 <
            N < 1 \times 10^7$ $^{31}$P spins in an InP NW.}  (a)
          In-phase $X(t)$ (black) and quadrature $Y(t)$ (light gray)
          force on the cantilever demodulated at the cantilever
          frequency.  $T = 4.2$ K and $B_{\text{ext}} = 6$ T.  The
          lock-in time constant is $\tau_l = 5$ s, in order to match
          the correlation time $\tau_m = 3.6$ s of the spin
          fluctuations and to reject the cantilever's thermal
          fluctuations with a correlation time $\tau_c = 65$ ms.  The
          thermal noise $Y(t)$ sets the limit for the MRFM detection
          sensitivity at a polarization equivalent to $\sim 250$
          $^{31}$P nuclear spins rms.  (b) Histograms of $X(t)$ and
          $Y(t)$ recorded for 1 hour.  Gaussian fits to these
          histograms give $\sigma_X = 6.4$ aN and $\sigma_Y = 2.1$
          aN.}
   \label{fig2}
\end{figure}

\begin{figure}[t]
	\includegraphics[width=\textwidth]{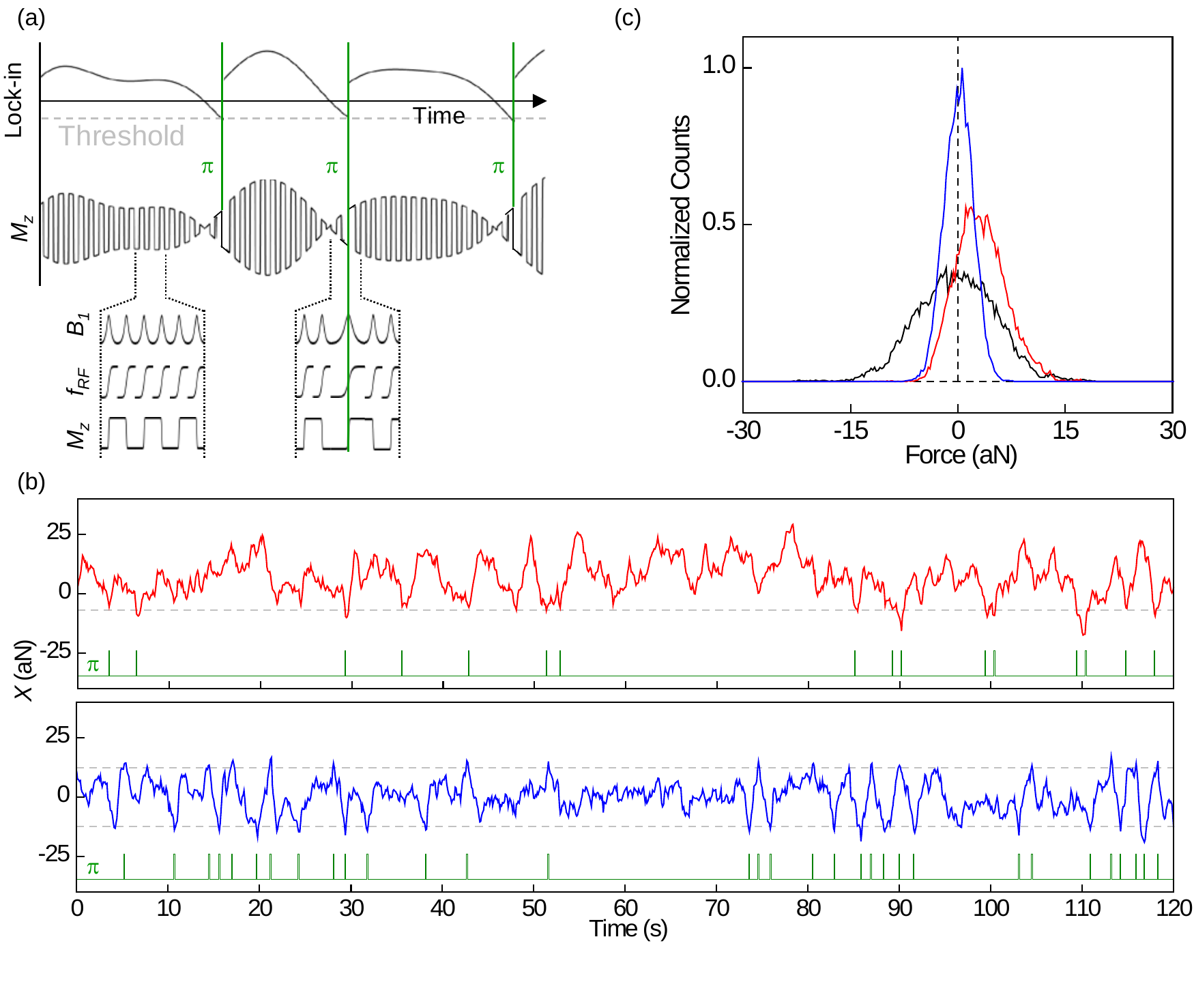}
	\caption{{\bf Rectifying and narrowing the nuclear spin
            polarizations.}  (a) Schematic diagram describing the
          conditional application of $\pi$-inversion pulses. (b)
          $X(t)$ recorded with a rectification threshold of -5 aN
          (top) and with an absolute value threshold of 12.5 aN
          (bottom).  Green pulses show the times at which
          $\pi$-inversions are applied.  $\tau_l = 0.8$ s in both
          cases.  The sample is an ensemble of $6 \times 10^5 < N < 1
          \times 10^7$ $^{31}$P spins in an InP NW at $T = 4.2$ K and
          $B_{\text{ext}} = 6$ T.  (c) Histograms of $X(t)$ recorded
          over 1 hour corresponding to the natural (black), rectified
          (red), and narrowed (blue) cases.  The mean polarization of
          the rectified distribution is $2.9$ aN compared with $0.0$
          aN of the natural distribution.  The standard deviation of
          the narrowed distribution is $2.0$ aN, compared with $5.6$
          aN of the natural distribution.}
   \label{fig3}
\end{figure}

\begin{figure}[t]
        \includegraphics[]{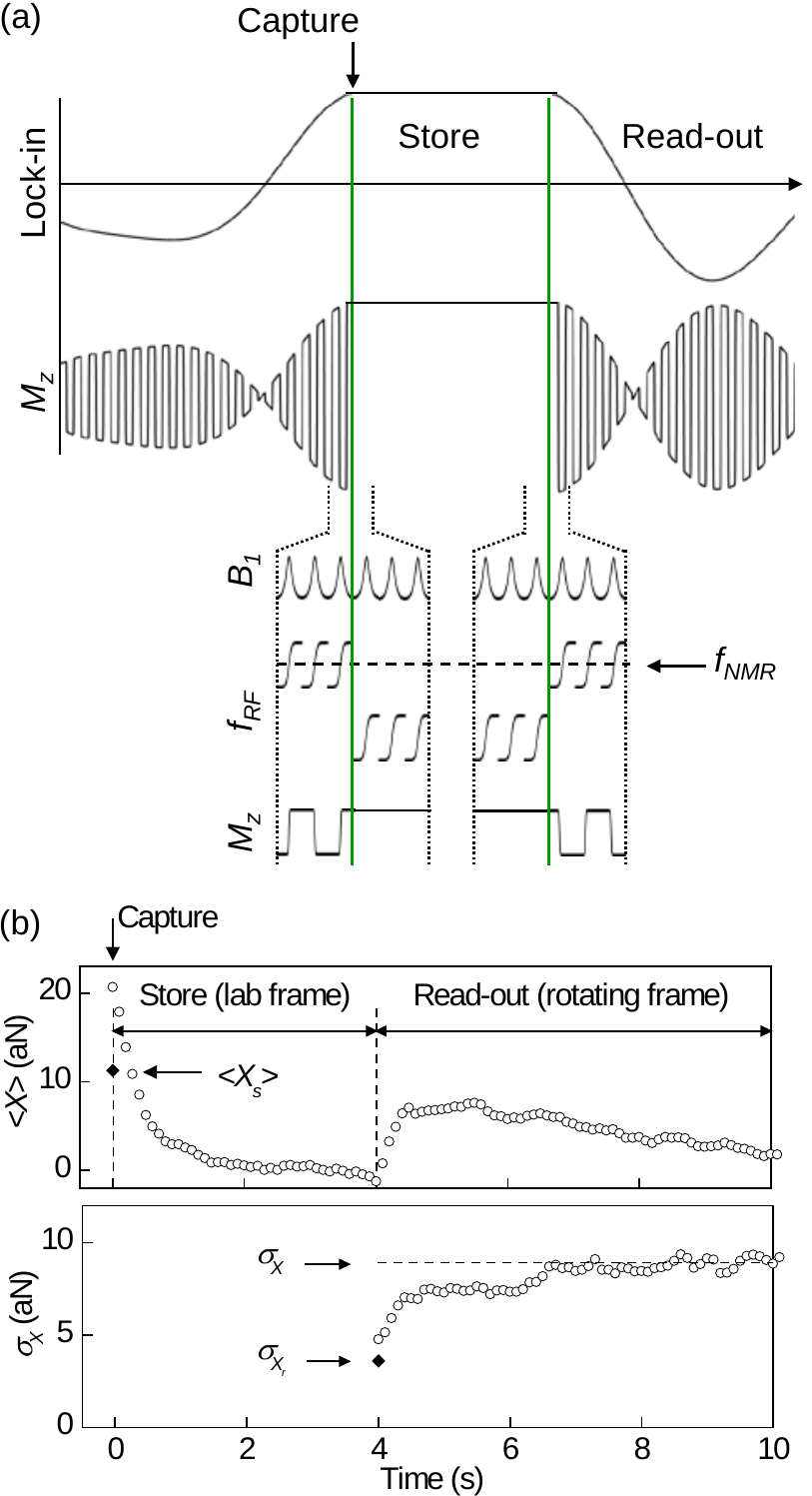}
        \caption{{\bf Capture-store-readout pulse sequence.}  (a)
          Schematic diagram describing the capture-store-readout pulse
          sequence.  (b) Top: $\left < X(t) \right >$ averaged over
          100 capture-store-readout sequences showing
          $T_{\text{store}} = 4$ s for a $^{31}$P spin ensemble with
          $6 \times 10^5 < N < 1 \times 10^7$ $^{31}$ in a GaP NW.
          The mean stored fluctuation $\left < X_s \right >$ is shown
          as a filled diamond. $T = 4.2$ K and $B_{\text{ext}} = 6$ T.
          The signal's decay after the capture is due to the lock-in
          time constant $\tau_l = 0.4$ s.  The read-out fluctuations
          decay over a time $\tau_m = 3.8$ s.  Bottom: $\sigma_X(t)$
          averaged over the same 100 sequences showing the narrowed
          variance of the retrieved fluctuation $\sigma_{X_r}$ (filled
          diamond) as it equilibrates back to its rotating frame value
          $\sigma_X$ (dotted line).}
  \label{fig4}
\end{figure}

\begin{figure}[t]
	\includegraphics[width=\textwidth]{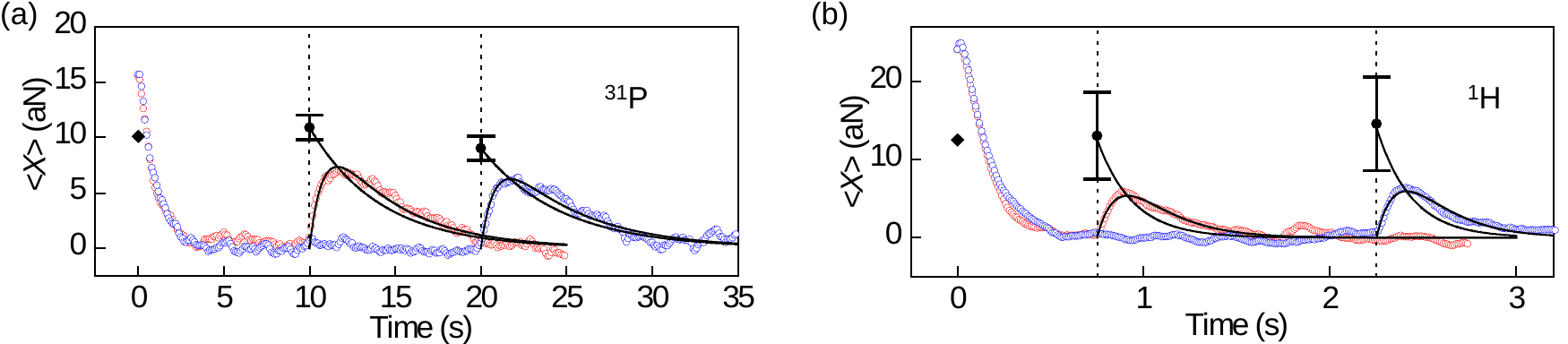}
	\caption{{\bf Storage of spin order in the laboratory frame.}
          (a) Two capture-store-readout sequences are shown for
          different $T_{\text{store}}$ for an ensemble of $6 \times
          10^5 < N < 1 \times 10^7$ $^{31}$P spins in an InP NW.  Fits
          to $X(t)$ during the readout take into account the lock-in
          time constant $\tau_l = 0.8$ s (solid lines) and allow us to
          recover the value of the retrieved fluctuation $\left < X_r
          \right >$ (filled circles) and its exponential decay with
          $\tau_m = 4.3$ s without the effect of $\tau_l$ (solid
          lines).  The filled diamond indicates the mean stored
          fluctuation $\left < X_s \right >$.  (b) The same
          measurement done for $2 \times 10^5 < N < 7 \times 10^5$
          $^1$H spins on a GaP NW with $\tau_l = 140$ ms where
          $\tau_m$ is found to be $190$ ms.  Again $\left < X_r \right
          >$ are displayed as filled circles and $\left < X_s \right
          >$ as a filled diamond.  In both cases $T = 4.2$ K and
          $B_{\text{ext}} = 6$ T.}
   \label{fig5}
\end{figure}

The MRFM consists of an nanomagnetic tip integrated on top of a
microwire RF source, an ultrasensitive Si cantilever, and a
fiber-optic interferometer to measure its displacement
\cite{Poggio:2007}.  The entire apparatus operates in vacuum better
than $10^{-6}$ mbar, at temperatures down to $T = 4.2$ K, and in an
applied longitudinal field up to $B_{\text{ext}} = 6$ T.  The NW of
interest is attached to the end of the cantilever and is positioned
within 100 nm of the nanomagnetic tip, as shown schematically in
Fig.~\ref{fig1}.  Here the field produced by the tip,
$\mathbf{B}_{\text{tip}}$, results in a total static magnetic field
$\mathbf{B}_0 = \mathbf{B}_{\text{ext}} + \mathbf{B}_{\text{tip}}$,
while the microwire produces a transverse RF field $\mathbf{B}_1(t)$.
ARP pulses invert the nuclear spin polarization because the
polarization follows (or is ``spin-locked'' to) the time-dependent
effective field $\mathbf{B}_{\text{eff}}(t) = \left ( B_0 - \frac{2
    \pi f_{\text{RF}}(t)}{\gamma} \right ) \mathbf{e}_z + \frac{1}{2}
B_1(t) \mathbf{e}_x$ in a frame rotating with the RF field, where
$f_{\text{RF}}(t)$ is the instantaneous frequency of the ARP pulses,
$B_1(t)$ their amplitude, and the unit vectors are defined in the
rotating frame.

The volume of inverted spins, known as the ``resonant slice'', is
determined by the spatial dependence of $\mathbf{B}_{\text{tip}}$ and
the parameters of the pulses.  This force variance is in turn related
to the number of nuclear spins in this volume by the approximation,
$\sigma_{S}^2 \approx N \frac{I(I+1)}{3} ( \hbar \gamma )^2 \left
  (\partial B_0 / \partial x \right)^2$, where $\mathbf{e}_x$ is the
direction of cantilever oscillation \cite{Xue:2011}.  This relation
holds as long as the volume of spins is small enough that the gradient
$\partial B_0 / \partial x$ is nearly constant throughout; otherwise
it provides a lower bound $N_{\text{lower}}$ on the number of spins.
From measurements of the magnetic field gradient in the vicinity of
the tip, we estimate that $\partial B_0 / \partial x = 1.5 \times
10^6$ T/m at the position of the detection volume (see supplementary
material).  In order to set an upper bound for the number of spins
$N_{\text{upper}}$, we make a model of the magnetic field profile
produced by the nanomagnetic tip and numerically integrate the
spatially dependent gradient over the sample volume contained in the
``resonant slice'' (see supplementary material).

Ultrasensitive cantilevers are made from undoped single-crystal Si and
measure 120 $\mu$m in length, 4 $\mu$m in width, and 0.1 $\mu$m in
thickness.  In vacuum and at the operating temperatures, the NW-loaded
cantilevers have resonant frequencies $f_c = 2.4$ and $3.5$ kHz,
intrinsic quality factors $Q_0 = 3.0 \times 10^4$ and $3.5 \times
10^4$, and spring constants $k = 60$ and $100$ $\mu$N/m for the InP
and GaP NW experiments respectively.  The InP NW is 8 $\mu$m-long; its
diameter shrinks from 200 nm to 60 nm along its length; and it is
tipped by a 60 nm diameter Au catalyst particle, left over from the
growth process.  The GaP NW is 10 $\mu$m-long; its diameter is 1.0
$\mu$m; and it has a 1.5-$\mu$m-long tapered tip which reaches 90 nm
in diameter at the Au droplet.  In this case we remove the Au droplet
by cutting of the end of the NW with a focused ion beam, resulting in
a 300-nm-diameter GaP tip.  Finally, we sputter a 5 nm layer of Pt
onto the NW in order to shield electrostatic interactions
\cite{Stipe:2001}.  Each NW is affixed to the end of the cantilever
with less than 100 fL of epoxy (Gatan G1).  In the attachment process,
we employ an optical microscope equipped with a long working distance
and a pair of micromanipulators.  During the measurement, the
cantilever is actively damped in order to give it a fast response time
of $\tau_c = 65$ ms.  Up to 50 nW of laser light at 1550 nm are
incident on the cantilever as part of the fiber-optic interferometer.
The nanomagnetic tips are truncated cones of Dy fabricated by optical
lithography \cite{Mamin:2012}.  For the InP (GaP) NW experiment the
tip measures 225 (280) nm in height, 270 (250) nm in upper diameter,
and 380 (500) nm in lower diameter.  The RF source, on which the Dy
tip sits, is a 2-$\mu$m-long, 1-$\mu$m-wide, and 200-nm-thick Au
microwire.  By positioning each NW within 100 nm of the combined
structure, the detection volume can be exposed to spatial magnetic
field gradients exceeding $1.5 \times 10^6$ T/m and RF $B_1(t)$ fields
larger than 20 mT without significant changes in the experimental
operating temperature.  We use hyperbolic secant ARP pulses with
$\beta = 10$ and a modulation amplitude set to 500 kHz peak-to-peak
for $^{31}$P and 1 MHz for $^1$H \cite{Tannus:1996}.  These
parameters, combined with the geometry of the sample and the profile
of $\partial B_0 / \partial x$, determine the size of the detection
volume \cite{Xue:2011}.

\begin{acknowledgements}

\section{Acknowledgments}

The authors thank C. L. Degen, C. Kl\"{o}ffel, T. Poggio, and
R. J. Warburton for illuminating discussions; Hari S. Solanki for
experimental assistance; and S. Keerthana for assistance with a
figure.  We acknowledge support from the Canton Aargau, the Swiss
National Science Foundation (SNF, Grant No. 200020-140478), the Swiss
Nanoscience Institute, and the National Center of Competence in
Research for Quantum Science and Technology.

\section{Author contributions}
  
P.P. and M.P conceived and planned the experiments in collaboration
with F.X. P.P. carried out the experiments. P.P. and M.P. analyzed the
data and wrote the manuscript. P.P. and F. X. prepared the samples and
devices. The nanowires were grown by H.I.T.H., S.A. and E.P.A.M.B. All
authors discussed the results and contributed to the manuscript.

\end{acknowledgements}

\end{document}